\documentstyle[amsmath,amssymb,11pt,epsfig]{article}
\begin{document}
\title{\bf Searching for New Physics in Leptonic Decays
of Bottomonium
\thanks{Research partially supported by CICYT under grant AEN99-0692}}
\author{Miguel Angel Sanchis-Lozano$^{a,b}$\thanks{Email: Miguel.Angel.Sanchis@uv.es}
\vspace{0.4cm}\\
(a) Instituto de F\'{\i}sica
Corpuscular (IFIC), Centro Mixto Universidad de Valencia-CSIC \\
(b) Departamento de F\'{\i}sica Te\'orica, Universidad de Valencia \\
Dr. Moliner 50, E-46100 Burjassot, Valencia (Spain)}
\date{}
\maketitle
\begin{abstract}
        New Physics can show up in various
        well-known processes already studied in the Standard Model, 
        in particular by modifying decay rates to some extent. In this work   
        I examine leptonic decays of $\Upsilon$ resonances
        of bottomonium below $B\bar{B}$ production, subsequent to  
        a magnetic dipole radiative structural transition of the 
        vector resonance yielding a
        pseudoscalar continuum state, searching for the existence 
        of a light Higgs-like neutral boson that would imply a slight 
        but experimentally measurable breaking of lepton universality.
\end{abstract}
\vspace{-12.2cm}
\large{
\begin{flushright}
  IFIC/02-23\\
  FTUV-02-0617\\
  hep-ph/0206156
  \date
\end{flushright} }
\vspace{13cm}
\begin{small}
\end{small}
\newpage

\section{Introduction}

The possible existence of Higgs-like bosons (hereafter referred as
$\lq\lq$Higgs'') beyond the Standard Model (SM) 
has already a long history in modern physics. For 
instance, let us mention the two-Higgs-doublet model (2HDM) \cite{gunion} 
- including the minimal supersymmetric
model as one of its specific realizations -  allowing for
five scalar or pseudoscalar bosons, three of which neutral and, perhaps,
one of them at least quite light. Another example of this kind is the axion
model, originally introduced in \cite{wilczek,weinberg} as a 
consequence of the spontaneous breaking of the global $U(1)_{PQ}$ 
axial symmetry. Nowadays, the
axion is a pseudoscalar field
appearing in a variety of theories with different meanings including
superstring theory, yielding
sometimes a massless particle and others a massive one: in
astrophysics the axion represents a good candidate of
the cold dark matter component of the Universe. In fact, there are other
scenarios, e.g. spontaneously broken lepton-number symmetry
\cite{gelmini,arcadi}, leading to
very light or even massless
bosons (Majorons) which might be detected in very
rare decays of strange and beauty particles
\cite{mas89}. On the other hand, another kind
of (neutral) scalar particles (familons) with either flavor-changing
or flavor-conserving couplings to fermions,
arise assuming a global family symmetry
spontaneously broken \cite{wilczek82}.
To end this quick outlook beyond the SM, a scalar particle (the radion)
proposed in the framework of 
the Randall-Sundrum model \cite{randall} of extra dimensions,
corresponding to the quantum excitations of
the interbrane separation \cite{gold}, could mix
with the standard Higgs and couple to SM fields \cite{rizzo}. Also
Kaluza-Klein (scalar) gravitons in an ADD scenario \cite{arkani,anto} could
eventually lead to measurable deviations from the SM \cite{han,giudice}.

Although there are well established lower mass 
bounds for the standard Higgs 
(e.g. from LEP searches \cite{lep}), the situation
may be different in several scenarios and models beyond the SM  
where such constraints would not
apply, leaving still room for
light Higgs bosons (see \cite{opal,haber01,maria} for example).
Needless to say, 
any possible experimental signal or discovery strategy
of Higgs-like particles should be examined with
great attention.

In this regard, let us remind that
the search for axions or light Higgs in the decays of 
heavy resonances has several attractive features. 
Firstly, the couplings of the former to fermions are 
proportional to their masses and therefore enhanced with respect to 
lighter mesons. Second, theoretical predictions are more reliable, 
especially with the recent development of effective theories like
non relativistic quantum 
chromodynamics (NRQCD) \cite{bodwin}, appropriate 
to deal with such bound states from first principles.
Indeed, intensive searches for a light Higgs-like boson (to be generically
denoted by $\phi^0$ in this paper)  have been
performed so far according to the so-called Wilczek mechanism \cite{wilczek2}
in the radiative decay of vector heavy quarkonia like
the Upsilon resonance (i.e. $\Upsilon{\rightarrow}\ \gamma \phi^0$). So far,
none of all these searches has been successful, but have 
provided valuable constraints on the mass values of light Higgs 
bosons \cite{gunion}.

Nevertheless, in this Letter I will focus on a possible signal
of New Physics based on the $\lq\lq$apparent'' breaking of  
lepton universality in 
bottomonium decays: {\em stricto sensu}, lepton universality
implies that the electroweak couplings 
to gauge fields of all charged 
lepton species should be the same. According to the interpretation
given in this work, the possible dependence
on the leptonic mass of the leptonic branching fractions 
of $\Upsilon$ resonances below the $B\bar{B}$ threshold (if 
experimentally confirmed by forthcoming measurements) might be viewed
as a hint of the existence of a Higgs of mass about 10 GeV.

\section{Searching for a light Higgs-like boson in $\Upsilon$ leptonic decays}

Let us get started by writing the well-known Van Royen-Weisskopf formula 
\cite{royen} including the color factor \footnote{As is well known, gluon
exchange in the short range part of the quark-antiquark potential
makes significant corrections to Eq.(1) but
without relevant consequences in our later discussion.} for  
the leptonic width of a vector quarkonium state 
without neglecting leptonic masses, 
\begin{equation}
{\Gamma}^0_{\ell\ell}\ =\ 
4\alpha^2Q_b^2\ \frac{|R_n(0)|^2}{M_{\Upsilon}^2}\ {\times}\ 
K(x)
\end{equation}
where $\alpha\ {\simeq}\ 1/137$ is the electromagnetic fine 
structure constant;
$M_{\Upsilon}$ denotes the mass of the vector particle 
(a $\Upsilon(nS)$ resonance in this particular case) and
$Q_b$ is the charge of the relevant (bottom) quark ($1/3$ in units of $e$);
$R_n(0)$ stands for the non-relativistic radial wave function of the
$b\overline{b}$ bound state at the origin; finally, the
$\lq\lq$kinematic'' factor $K$ reads
\begin{equation}
K(x)\ =\ (1+2x)(1-4x)^{1/2}
\end{equation}
where $x=m_{\ell}^2/M_{\Upsilon}^2$.
Leptonic masses are usually neglected in Eq.(1) (by setting $K$
equal to unity) except for the decay into $\tau^+$ $\tau^-$ 
pairs. Let us note that $K(x)$ is a decreasing function of $x$:
the higher leptonic mass the smaller decay rate. However, 
such $x$-dependence is quite weak for bottomonium.

In this work, I am conjecturing the existence of a light Higgs-like 
particle whose mass is close
to the $\Upsilon$ mass and which could show up in the sequential decay:
\begin{equation}
\Upsilon\ {\rightarrow}\ \gamma\ \phi^0({\rightarrow}\ \ell^+\ell^-)\ \ \ ;\ 
\ \ \ell=e,\mu,\tau
\end{equation}
Actually, this process may be seen
as a continuum radiative transition that in principle permits the
coupling of the bottom quark-antiquark pair to a particle 
of variable mass and $J^{PC}$:
$0^{++}, 0^{-+}, 1^{++}, 2^{++}...$ (always positive charge conjugation).
In the present investigation, we will confine our attention to the two 
first possibilities: a scalar or a pseudoscalar boson.

Let us remark that if indeed the $\phi^0$ mass were quite close to 
the $\Upsilon$ mass, on the one hand the 
(pseudo)scalar propagator could enhance the width of process (3);
on the other hand, the emitted photon would  become soft enough to remain 
experimentally below detection threshold and the 
contribution (3) would be experimentally ascribed to the leptonic channel, 
thereby (slightly) increasing its decay rate.
Both related points constitute cornerstones along this paper
and will be discussed in more detail.

\subsection{An intermediate spin-singlet $b\bar{b}$ state?}

Motivated by the study of heavy quarkonia production and
decay, where intermediate non-perturbative states
are thought to play an important (sometimes 
leading) role \cite{bodwin}, 
one may envisage the existence of equivalent intermediate hadronic stages 
after the radiative decay of a vector resonance as
shown in Eq.(3), before
annihilating into a charged lepton pair \footnote{I leave aside 
the discussion on the
possible formation of the ${\eta}_b$ state 
(still unobserved at present except for a recent claim
\cite{aleph02}): in such a case
only the  $0^{-+}$ state and a sharply defined mass would be allowed.
Nevertheless, notice that the foreseen mass difference 
between the $\Upsilon(1S)$ and ${\eta}_b$ states,
recently estimated in \cite{vairo} to be 36-55 MeV, makes in practice
almost no difference with respect to our later analysis based
on a continuum state.}.

In a naive quark model, quarkonium is treated as a nonrelativistic bound
state of a quark-antiquark pair in a static color field which sets up  
an instantaneous confining potential. Indeed this picture has been 
remarkably successful in accounting for the properties
of heavy quarkonia. However, it overlooks gluons whose wavelengths are
larger that the bound state size: dynamical gluons permit
a Fock decomposition of physical quarkonium states
beyond the leading non-relativistic description.

Thus, in a vector resonance like the $\Upsilon(1S)$, the heavy quark 
pair can be in a spin-trilet, color-singlet state in the lowest 
Fock state, but the $b\bar{b}$ system could also exist in 
a configuration other than $J^P=1^-$ since the soft degrees of freedom
can carry the remaining quantum numbers, although with a
smaller probability. Chromoelectric and chromomagnetic transitions 
can connect different Fock states in physical processes
subject to selection rules and governed by distinct
probabilities.
These ideas have been cast into the rigorous
framework of an effective non-relativistic theory for the strong interaction 
(NRQCD) \cite{bodwin}, widely applied 
to heavy quarkonia phenomenology (see, for example, \cite{mas97,mas01}
and references therein). Furthermore, the authors of \cite{wolf} have shown
the physical relevance of soft gluon emission in the
(semi)leptonic decays of the $B_c$ meson.
Let us finally mention that the possibility of intermediate
bound states in electromagnetic decays of quarkonia in particular and its
physical consequences have 
been discussed elsewhere \cite{efetov}, without resorting
to a Fock decomposition.

At this point, one may wonder about the possibility of reaching different
$b\bar{b}$ bound states starting from an initial 
spin-triplet color-singlet configuration
by radiation of soft photons, instead of gluons, via electric and magnetic
transitions. 
However, the situation is not quite the same as in $\lq\lq$standard''
NRQCD. In effect, let us emphasize one of
consequences of the the crucial difference between photons 
and gluons: since the latter carry color,  
a lower energy bound always applies in soft emission
at the final stage hadronization 
(corresponding, for instance, to a pion mass). On the contrary, 
no such infrared cut-off 
exists for photons, except for the
experimental detection threshold which properly regularizes the infrared
divergences \cite{jauch}. 
Moreover, (soft) single gluon emission from quarkonia would leave the heavy
quark-antiquark pair in a color-octet state, thereby forbidding 
its subsequent annihilation into
a charged lepton pair. Therefore, soft photon emission in leptonic decays 
of $\Upsilon$ resonances can have important consequences
beyond purely radiative corrections.

In fact, several authors \cite{haber79,ellis79} suggested long time ago
that a scalar ($h^0$)  Higgs 
of mass near 10 GeV could enhance the tauonic decay of
the $\chi_{b0}$ resonance, proposing a close look at the cascade
decay with an $e^+e^-$ machine working on the $\Upsilon(2S)$ resonance:
\begin{equation}
\Upsilon(2S){\rightarrow}\ \gamma\ \chi_{b0}(\rightarrow h^0{\rightarrow}\ 
\tau^+\tau^-)
\end{equation}
where the radiative decay corresponds (mainly) to an electric dipole 
(E1) transition.
Actually, our underlying idea is not too far from 
theirs, although now the structural magnetic dipole (M1) transition should 
no lead to a P-wave resonance, 
but likely to a S-wave pseudoscalar continuum bound state. 
Indeed, let us note that a E1 transition would
involve the product of a $S$-wave initial-state function, peaking at the
origin, with a $P$-wave final-state function, vanishing at the origin; 
in the long wavelength limit
the overlap should be small. Besides, it seems unrealistic
a continuum P-wave $b\bar{b}$ state whose mass is smaller than the initial
S-wave state mass. The last argument can be extended to other
quantum numbers of the continuum resonance and, consequently, 
I dismiss all possibilities but a pseudoscalar
$b\bar{b}$ bound state after photon emission in the process (3).

\begin{figure}[htb]
\centerline{\hbox{
 \psfig{file=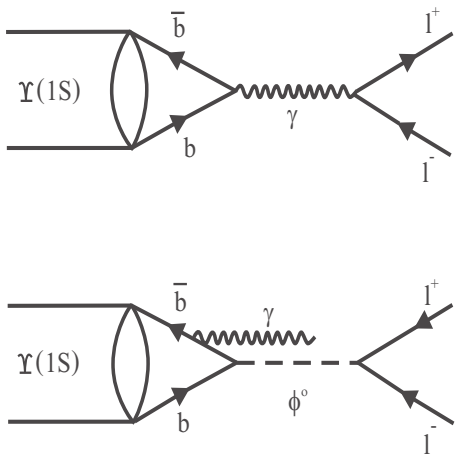,height=6.cm,width=6cm}  
}}
\caption{(a)[upper panel]: Electromagnetic 
annihilation of a $b\bar{b}[{}^3\!S_1^{(1)}]$ 
bound state (a $\Upsilon(1S)$ resonance
in particular) into a charged
lepton pair through a vector particle (i.e. a photon);
(b)[lower panel]: Hypothetical annihilation of a color-singlet 
$b\bar{b}{[{}^1\!S_0^{(1)}]}$ state (assuming that the
$b\bar{b}{[{}^3\!S_1^{(1)}]}$ resonance state
has previously undergone a M1 transition into a pseudoscalar
continuum state, caused by soft photon emission) 
into a charged lepton pair through a pseudoscalar Higgs-like boson 
(represented  
by a dashed line in the figure and denoted by $\phi^0$ in the text).}
\end{figure}

In sum, magnetic dipole transitions of
the $\Upsilon$ into a non-perturbative
$b\bar{b}[{}^1\!S_0^{(1)}]$ state 
(I am using spectroscopic notation and stressing its
color-singlet status by keeping the superscript $(1)$ widely
used in NRQCD) can play a noticeable
role in a process like (3).
The corresponding width can be estimated with the aid of
a text-book formula \cite{oliver}:
\begin{equation}
{\Gamma}^{M1}_{\Upsilon(1S){\rightarrow}\gamma_sb\bar{b}[{}^1\!S_0^{(1)}]}\  
{\simeq}\ \frac{16 \alpha}{3}\biggl(\frac{Q_b}{2m_b}\biggr)^2k^3
\end{equation}
where $k$ denotes the energy of the soft photon $\gamma_s$, radiated
from either the heavy quark or the heavy antiquark inside 
bottomonium, as shown in 
figure 1.b; the mass of the Upsilon resonance will be approximately taken as
twice the bottom quark mass throughout the
paper, i.e. $M_{\Upsilon} \simeq 2m_b$.
Actually, the above equation should provide just an 
order-of-magnitude estimate   
since, for instance, the density of the
continuum $b\bar{b}{[{}^1\!S_0^{(1)}]}$
states is not taken into account
\footnote{One could improve somewhat Eq.(5) by means of 
a multiplicative overlap factor of
the wave functions of both resonances as in
\cite{close}. Note however that for closely spaced
states this overlap factor should be near unity. See also
Ref. \cite{wong} for a thorough calculation of chromoelectric and
chromomagnetic transitions in heavy quarkonium.}; but, in reality, the main 
uncertainty comes from the value used for the soft 
$\lq\lq$typical'' photon energy, i.e. the input $k$ in Eq.(5).

In this regard, note that typical widths 
of resonance peaks in $e^+e^-$ machines
are of the order of one or few tens of MeV for the $\Upsilon$
family below open bottom production (much
larger than resonance widths themselves). On the other hand, usual
thresholds for photon detection are of order 50 MeV. Moreover, 
in the case of tauonic decays, event selection is based on 
either semileptonic or hadronic inclusive
decays of the $\tau$'s (usually requiring a $\lq\lq$back-to-back'' topology) 
with similar (probably larger) uncertainties as in the electron or 
muonic channels, concerning a soft
photon escaping detection \cite{cleo94}.

Therefore, letting the photon energy $k$ vary over the range 
$10-50$ MeV, one gets from (5) the interval 
$\Gamma^{M1}_{\Upsilon(1S){\rightarrow}\gamma_sb\bar{b}[{}^1\!S_0^{(1)}]}\ 
{\simeq}\ 4.3{\times}10^{-5}-5.4{\times}10^{-3}$ keV. 
Let us stress that those small $k$ values
do justify the use of Eq.(5) for magnetic dipole transitions since 
the wavelengths of the radiated photons are
quite larger than the size of quarkonium 
(of order $\simeq {\cal O}(1)$ GeV$^{-1}$). 

Next, dividing the partial width given in Eq.(5)
by the total width of
the resonance, $\Gamma_{tot}=52.5$ keV \cite{pdg}, one gets
the probability that the M1 transition takes place, i.e. 
\begin{equation}
{\cal P}^{M1}_{\Upsilon(1S){\rightarrow}\gamma_sb\bar{b}[{}^1\!S_0^{(1)}]}\ 
=\ \frac{\Gamma^{M1}_{\Upsilon(1S)
{\rightarrow}\gamma_sb\bar{b}[{}^1\!S_0^{(1)}]}}
{\Gamma_{tot}}\ {\simeq}\ 
\frac{1}{\Gamma_{tot}}\ \frac{4\alpha Q_b^2}{3m_b^2}\ k^3 \ 
\simeq\ 8.2{\cdot}10^{-7}-10^{-4}
\end{equation}
The latter quantity can be interpreted as a kind of branching fraction (BF)
for the radiative decay of the $\Upsilon(1S)$ into the supposed
continuum $b\bar{b}$ spin-singlet resonance, under
the experimental constraint of the photon energy upper cut-off.   
The range shown in Eq.(6) seems realistic when compared to the
measured BF for the channel $J/\psi{\rightarrow}\gamma\eta_c$,
(of order $1\%$ \cite{pdg}) taking into account the different mass and
electric charge for the charm and bottom quarks.

Subsequently, the pseudoscalar $b\bar{b}[{}^1\!S_0^{(1)}]$ 
state could decay into a 
charged lepton pair at tree-level
via the conjectured neutral boson, as will be discussed in more
detail in the next section.

Finally let us emphasize that, in this experimental context, soft photons 
(mainly radiated by the initial-state  
colliding electrons in $e^+e^-$ machines and by  the
final-state outgoing leptons) are those
whose energies do not exceed the experimental detection
threshold and thereby
can not be observed \footnote{In the language of low-energy effective theories
such photons would be properly
designed as ultrasoft \cite{labelle,soto}.}.
Nevertheless, as is well-known soft radiation corrections must be included
in the theoretical analysis \cite{kuraev,pdg} for a meaningful comparison with 
the experimental results for the leptonic branching fraction:
${\cal B}_{\ell\ell}={\Gamma}_{\ell\ell}/{\Gamma}_{tot}$.
In an analogous fashion, photons from
the conjectured M1 transition of
the $\Upsilon$ resonance would merge with the aforementioned
bulk of soft radiation, without changing 
significantly the invariant mass of the lepton pair 
(i.e. within mass reconstruction windows), or
passing trigger and off-line
selection criteria, ultimately contributing to
the measured leptonic BF.

\subsection{Effects of a light neutral Higgs on the leptonic 
decay width} 

In this Letter, we are focusing on vector 
$\Upsilon(nS)$ states of the bottomonium family below
open flavor (i.e. $n<4$) \footnote{The $\Upsilon(3S)$ state is
excluded in the present analysis
since only experimental 
data available for the muonic channel \cite{pdg} are currently
available; see http://pdg.lbl.gov for regular updates.} decaying 
into a charged lepton pair plus a {\em soft} photon:
\begin{equation}
\Upsilon(nS){\rightarrow}\ \gamma_s\ b\bar{b}[{}^1\!S_0^{(1)}](\rightarrow 
\phi^0{\rightarrow}\ \ell^+\ell^-)\ \ \ ;\ 
\ \ \ell=e,\mu,\tau
\end{equation}
where the soft (I stress it once more: {\em unobserved}) 
$\gamma_s$ comes from a  
{\em prompt} M1 transition of the $\Upsilon$ resonance,
as sketched in figure 1.b.

The decay width ${\Gamma}_{{\gamma_s}\ell\ell}$ will be written
in a factored form reflecting the two-step process:
prior formation of an intermediate (mainly) pseudoscalar state of account of
soft magnetic radiation,
followed by its annihilation into a charged lepton pair
mediated by a (pseudo)scalar Higgs:
\begin{equation}
{\Gamma}_{\gamma_s\ell\ell}=
{\cal P}^{M1}_{\Upsilon(1S){\rightarrow}\gamma_sb\bar{b}[{}^1\!S_0^{(1)}]} 
\times \tilde{\Gamma}_{\ell\ell}
\end{equation}
where $\tilde{\Gamma}_{\ell\ell}$ is the annihilation width of
the $b\bar{b}[{}^1\!S_0^{(1)}]$ state
into a lepton pair via $\phi^0$ Higgs exchange.
Furthermore, fermions are supposed to couple to the
$\phi^0$ field
according to a Yukawa interaction term in the effective Lagrangian,
\begin{equation}
{\cal L}_{int}^{\bar{f}f}\ =\ -\xi_f^{\phi}\ \frac{\phi^0}{v}
m_f\bar{f}(i\gamma_5)f 
\end{equation}
where $v=246$ GeV stands for the vacuum expectation value 
of the standard Higgs boson; $\xi_f^{\phi}$ denotes a factor
depending on the type of the Higgs boson, which
could enhance the coupling with a fermion 
(quark or lepton) of type $\lq\lq$$f$''. In particular, $\phi^0$ 
couples to
the final-state leptons proportionally to their masses, 
ultimately required because
of spin-flip in the interaction of a fermion with a 
(pseudo)scalar, expectedly providing an experimental 
signature for checking the existence of a light Higgs in
our study.
Lastly, note that the
$i\gamma_5$ matrix stands only in the case of 
a pseudoscalar $\phi^0$ field.

In this paper, I am tentatively assuming that the mass of the sought
light Higgs stands 
close to the $\Upsilon(1)$ resonance but below $B\bar{B}$ production:
$m_{\phi^0}\ {\simeq}\ 2m_b$. As will be argued 
from current experimental data in the
next section, I suppose specifically that $m_{\phi^0}$ lies
somewhere between the $\Upsilon(1S)$ and $\Upsilon(2S)$ masses,
i.e.
\begin{equation}
m_{\Upsilon(1S)}\ {\lesssim}\ m_{\phi^0}\ {\lesssim}\ m_{\Upsilon(2S)}
\end{equation}

Now, I define the mass difference:
${\delta}m=|m_{\phi^0}-m_{\Upsilon}|$, where $\Upsilon$
denotes either a $1S$ or a $2S$ state. Accepting for simplicity that
the Higgs stands halfway between the mass values of both
resonances, I will set ${\delta}m\ {\simeq}\ 0.25$ GeV
for an order-of-magnitude calculation. Hence I write
approximately for the scalar tree-level propagator 
of the $\phi^0$ particle in (7),
\begin{equation}
\frac{1}{(m_{\Upsilon}^2-m_{\phi^0}^2)^2}\ \simeq\ 
\frac{1}{16\ m_b^2\ {\delta}m^2}
\end{equation}
where the width of the Higgs boson has been neglected 
\footnote{In case that the resonance and
the $\phi^0$  would stand closer than one Higgs width, 
there would be a mixing between both states further enhancing
the decay into lepton pairs \cite{ellis79,drees}, deserving
a separate study.} for it should
be very narrow due the smallness of $m_{\phi^0}$ and,
moreover, standing below bottom open production according to 
the inequality (10). 

Performing a comparison between the widths of 
both leptonic decay processes
(i.e. $\tilde{\Gamma}_{\ell\ell}$ versus
${\Gamma}_{\ell\ell}^0$), one concludes with the aid of Eqs.(9-11)
and (1) that 
\begin{equation}
\tilde{\Gamma}_{\ell\ell}\ \approx\  
\frac{3m_b^4m_{\ell}^2K(x)|R_n(0)|^2\xi_b^2\xi_{\ell}^2}
{2\pi^2(m_{\Upsilon}^2-m_{\phi^0}^2)^2v^4}\ \simeq\ 
\frac{3\xi_b^2\xi_{\ell}^2}{32\pi^2Q_b^2\alpha^2}
\frac{m_b^4m_{\ell}^2}{{\delta}m^2v^4}\ {\times}\ 
\Gamma_{\ell\ell}^0
\end{equation}
with $K(x)$ defined in Eq.(2).

Above I used
a non-relativistic approximation
\footnote{More precisely, I used the static approximation for 
heavy quarks inside quarkonium (i.e. null relative momentum); then, the 
decay amplitude of the pseudoscalar state
into a  $0^{++}$ (CP-even) Higgs vanishes and only a $0^{-+}$ 
(CP-odd) Higgs couples to pseudoscalar quarkonium in this limit. 
Therefore, the Higgs hunted in this work
could be properly denoted by $A^0$ though I
will keep using the generic  $\phi^0$ symbol for it, though
standing hereafter for a CP-odd Higgs.}, assuming the same
wave function at the origin for both the $\Upsilon$ vector
state and the $b\bar{b}[{}^1\!S_0^{(1)}]$ intermediate bound state
on account of heavy-quark spin symmetry \cite{bodwin}.

Finally,
the BF for channel (7) can be readily obtained from Eq.(8),
with the help of Eqs. (5) and (12), as
\begin{equation}
{\cal B}_{\Upsilon{\rightarrow}\gamma_s\ell\ell}= 
\biggl[\frac{m_b^2m_{\ell}^2k^3\xi_b^2\xi_{\ell}^2}
{8\pi^2{\alpha}\Gamma_{tot}{\delta}m^2v^4}\biggr] 
\times {\cal B}_{\ell\ell}
\end{equation}
so one can compare
the relative rates by means of the following dimensionless ratio
\begin{equation}
{\cal R}=\frac{{\cal B}_{\Upsilon{\rightarrow}\gamma_s\ell\ell}}
{{\cal B}_{\ell\ell}}=
\biggl[\frac{m_b^2k^3\xi_b^2\xi_{\ell}^2}
{8\pi^2{\alpha}\Gamma_{tot}v^4}\biggr] 
\times \frac{m_{\ell}^2}{{\delta}m^2}
\end{equation}
where we are assuming that the main contribution to the leptonic
channel comes from the photon exchange graph of figure 1.a. 
Let us point out once again that since $\gamma_s$ is undetected, 
the Higgs contribution of figure 1.b would be experimentally ascribed
to the leptonic channel of the $\Upsilon$ resonance. 

For the sake of a comparison
with other Higgs searches, I will identify in the following
the $\xi_f$ factor with
the 2HDM (type II) parameter for the universal down-type
fermion coupling to a CP-odd Higgs, i.e. $\xi_b=\xi_{\ell}=\tan{\beta}$,
defined as the ratio of the vacuum expectation values of
two Higgs fields 
\cite{gunion}. Inserting
numerical values, one gets the interval
\begin{equation}
{\cal R}\ \simeq\ (3.6{\cdot}10^{-9}-4.5{\cdot}10^{-7})\ {\times}\ 
\tan{}^4\beta\ \times\ m_{\ell}^2
\end{equation}
where use was made of the approximation
$m_{\phi^0} \simeq 2m_b\ \simeq 10$ GeV, and 
the range $10-50$ MeV for the soft photon energy $k$; $m_{\ell}$
is expressed in GeV.

\section{Lepton universality breaking: hypothesis testing}

Now, I will confront the above predictions with the experimental results 
on $\Upsilon$ leptonic decays \cite{pdg} summarized in Table 1. 
(Notice that although those
BF's were often determined from raw data under the assumption
of lepton universality, its possible small breaking would not alter
considerably the final figures.) Indeed, from inspection of
Table 1 one realizes that experimental 
data favor a slight but steady increase of the decay rate
with the lepton mass in all cases. In spite of that, current 
error bars ($\sigma_{\ell}$) are still too large 
(especially in the case of the $\Upsilon(2S)$) to
permit a thorough check of the lepton mass dependence as
expressed in (15).

Nevertheless, I will apply below a hypothesis test
in order to draw, if possible, a statistically
significant conclusion about lepton 
universality breaking. To this end, I present in Table 2
the differences ${\Delta}_{\ell\ell'}$ between 
BF's of distinct channels obtained from Table 1. 
In order to treat  all of them in equal footing, such differences
are divided by their respective
experimental errors $\sigma_{\ell\ell'}$ (see Table 2).

\begin{table} [htb]
\caption{Measured leptonic BF's 
(${\cal B}_{\ell\ell}$) and error bars ($\sigma_{\ell}$) in $\%$, of
$\Upsilon(1S)$ and $\Upsilon(1S)$ (from \cite{pdg}).}
\begin{center}
\begin{tabular}{cccc}
\hline
channel: & $e^+e^-$ & $\mu^+\mu^-$ & $\tau^+\tau^-$  \\
\hline
$\Upsilon(1S)$ & $2.38 \pm 0.11$ & $2.48 \pm 0.06$ & $2.67 \pm 0.16$ \\
\hline
$\Upsilon(2S)$ & $1.18 \pm 0.20$ & $1.31 \pm 0.21$ & $1.7 \pm 1.6$ \\
\hline
\end{tabular}
\end{center}
\end{table}

\begin{table} [htb]
\caption{All six differences ${\Delta}_{\ell\ell'}$ 
(obtained from Table 1) between the leptonic BF's (in $\%$) of
$\Upsilon(1S)$ and $\Upsilon(2S)$ resonances separately.
Subscript ${\ell}\ell'$ denotes the difference between
channels into ${\ell}\bar{\ell}$ and ${\ell}'\bar{\ell}'$ 
lepton pairs
respectively, i.e.
${\Delta}_{\ell\ell'}={\cal B}_{\ell'\ell'}-{\cal B}_{\ell\ell}$; the
${\sigma}_{\ell\ell'}$  values were obtained from Table 1 by summing
the error bars in quadrature, i.e. 
$\sigma_{\ell\ell'}=\sqrt{\sigma_{\ell}^2+\sigma_{\ell'}^2}$. Only two 
${\Delta}_{\ell\ell'}/{\sigma}_{\ell\ell'}$ values for
each resonance can be considered as truly independent, amounting 
altogether to a total number of four independent points.}
\begin{center}
\begin{tabular}{cccc}
\hline
channels & ${\Delta}_{\ell\ell'}$ & $\sigma_{\ell\ell'}$ & 
${\Delta}_{\ell\ell'}/\sigma_{\ell\ell'}$ \\
\hline
$\Upsilon(1S)_{e{\mu}}$  & $0.1$ & $0.125$ & $+0.8$ \\
\hline
$\Upsilon(1S)_{\mu{\tau}}$  & $0.19$ & $0.17$ & $+1.12$ \\
\hline
$\Upsilon(1S)_{e{\tau}}$  & $0.29$ & $0.19$ & $+1.53$ \\
\hline
$\Upsilon(2S)_{e{\mu}}$  & $0.13$ & $0.29$ & $+0.45$ \\
\hline
$\Upsilon(2S)_{\mu{\tau}}$  & $0.39$ & $1.61$ & $+0.24$ \\
\hline
$\Upsilon(2S)_{e{\tau}}$  & $0.52$ & $1.61$ & $+0.32$ \\
\hline
\end{tabular}
\end{center}
\end{table}

On the other hand, we are especially interested 
in the {\em alternative} hypothesis 
based on the existence of a light Higgs boson
enhancing the decay rate as a growing function of
the leptonic squared mass, in opposition to the kinematic factor (2).
Therefore, the {\em region of rejection} of our
statistical test should lie only on {\em one side} (or {\em tail}) 
of the ${\Delta}_{\ell\ell'}/\sigma_{\ell\ell'}$ variable distribution
(i.e. positive values if $m_{\ell'}>m_{\ell}$), in particular
above a  preassigned {\em critical value} \cite{frodesen}.
In other words, I will perform
a {\em one-tailed test} \cite{frodesen} using the sample consisting
of four independent BF differences between 
the electronic and the muonic and tauonic channels respectively
(i.e. ${\Delta}_{e\mu}/\sigma_{e\mu}$,  ${\Delta}_{e\tau}/
\sigma_{e\tau}$) for both
$\Upsilon(1S)$ and $\Upsilon(2S)$ resonances shown in Table 2. 
I will assume that such differences follow a normal probability distribution. 
Then, the mean of the four ${\Delta}_{e\ell'}/{\sigma}_{e\ell'}$ values
($\ell'=\mu,\tau$) turns out to be $0.775$. 

Next, I define the {\em test statistic}: 
$T={\langle}{\Delta}_{e\ell'}/{\sigma}_{e\ell'}{\rangle}{\times}
\sqrt{N}=1.55$, where $N=4$ stands for the number of 
independent points. (Note that we are dealing with a 
Gaussian of unity variance after dividing all differences by
their respective errors.)
Now, choosing the {\em critical value} to be ${\simeq}\ 1.3$,  
the lepton universality hypothesis [playing the role  of the {\em null
hypothesis} in our test, predicting a mean zero (or slightly less) value]
can be {\em rejected} at 
a {\em significance level} of 10$\%$ since $T>1.3$ 
\footnote{Let us recall \cite{frodesen}
that in a (simple) hypothesis test,
the {\em significance level} (or {\em error of the first kind}) represents
the percentage of all decisions such that the null hypothesis
was rejected when it should, in fact, have been accepted.}.
Certainly, this result alone is
not statistically significant enough to make any serious claim 
about the rejection of the lepton universality hypothesis in
this particular process, but points out the interest to
investigate further the alternative hypothesis stemming from Eq.(15).

\section{Final results and discussion}

On the grounds of the hypothesis test carried out
in the previous section, I conclude
that current experimental data shown in Table 1 are
compatible with a (small) breaking of lepton universality 
in $\Upsilon$ leptonic decays at a significance level of $10\%$.
Indeed, there seems to be a slight but measurable
increase of the leptonic decay rate, 
by a ${\cal O}(10)\%$ factor, from the electronic channel to the 
tauonic channel, which can be interpreted theoretically following the
2HDM upon a reasonable choice of its parameters. 

Actually, from Eq.(15) one obtains that, under the assumption of a
Higgs boson of mass about 10 GeV to explain such a 
10$\%$ enhancement, $\tan{\beta}$ should roughly lie over the range:
\begin{equation}
16\ {\lesssim}\ \tan{\beta}\ {\lesssim}\ 54
\end{equation}
depending on the value of $k$, namely from 50 MeV to 10 MeV, whose limits
remain somewhat arbitrary however. Thus a 
caveat is in order: the above interval is purely indicative as it
only takes into account the probability range on the M1 transition 
estimated according to Eq.(5), and not other sources
of uncertainty. In this regard I have not considered, for example, the
possible mixing of the CP-odd Higgs with the pseudoscalar
$b\bar{b}$ system stemming from the $\Upsilon$ after photon radiation, which
nevertheless could modify notably the properties of the former 
\cite{drees}.

It is worthwhile to remark that the $\tan{\beta}$ interval of Eq.(16) 
is compatible with the range needed
to interpret the $g-2$ muon anomaly in terms of a
light CP-odd Higgs ($A^0$) resulting from a   
two-loop calculation \cite{maria,cheung}
\footnote{However, after correcting a sign mistake
in the so-called hadronic light by light contribution
in the $g-2$ calculation 
\cite{knecht,yndurain2} the discrepancy with
respect to the SM becomes smaller than initially 
expected. Still the situation is 
under close scrutiny (see, for example, \cite{rafael}).}. 
In this regard, other authors considered 
a CP-even Higgs ($h^0$) in a one-loop calculation
\cite{haber01} also yielding parametric values very
close to ours; nevertheless, note that the possibility of 
a scalar Higgs, contributing 
to the $\Upsilon$ leptonic decay as conjectured in this work,
was discarded in our study since the corresponding partial
width vanishes in the
(leading) static approximation for the $b\bar{b}[{}^1\!S_0^{(1)}]$ 
intermediate state. 

In summary, I have pointed out in this Letter  
a possible breaking of lepton universality 
in $\Upsilon$ leptonic decays, interpreted in terms of 
a neutral CP-odd Higgs of mass around $10$ GeV, introducing a $m_{\ell}^2$
dependent contribution in the partial width. (Notice that higher-order
corrections within the SM can not imply such a quadratic 
dependence on the leptonic mass.) 
Pictorially, one could say that the Higgs is $\lq\lq$hidden'' in part
because the (undetected) M1 photon, stemming from the assumed prompt 
structural
transition, merges with the bulk of soft photon radiation
incorporated into the very definition of the leptonic BF \cite{pdg}.  

I end by emphasizing the interest in more accurate data on
leptonic BF's of $\Upsilon$ resonances, 
particularly considering the exciting possibility of a signal
of New Physics as pointed out in this work.
Hopefully, B factories working below open bottom production will
provide in a near future new and likely more precise measurements of the 
leptonic BF's for the $\Upsilon$
family, perhaps requiring specific dedicated data-taking periods
if a light CP-odd Higgs discovery strategy would be
conducted including a complementary search for a CP-even Higgs
in accordance with the process (4). 

\subsection*{Acknowledgements} 
I thank F.J. Botella, J.J. Hern\'andez, P.Gonz\'alez, F. Mart\'{\i}nez-Vidal,
J. Papavassilou, A. Santamaria, J. A. Valls and S. Wolf 
for helpful discussions.

\end{document}